\documentclass[conference]{IEEEtran}
\IEEEoverridecommandlockouts
\usepackage{cite}
\usepackage{amsmath,amssymb,amsfonts}
\usepackage{algorithmic}
\usepackage{graphicx}
\usepackage{textcomp}
\usepackage[colorlinks=true, allcolors=blue]{hyperref}
\usepackage{xcolor}
\begin{document}

\title{Prefetcher-based DRAM Architecture\\
%\large{CS531 - Memory System \& Architecture}

}

\author{\IEEEauthorblockN{ Saurabh Jaiswal}
\IEEEauthorblockA{\textit{Computer Science and Engineering} \\
\textit{Indian Institute of Technology}\\
Ropar, India \\
saurabh.19csz0009@iitrpr.ac.in}
\and
\IEEEauthorblockN{ Shailendra Kumar Gupta}
\IEEEauthorblockA{\textit{Computer Science and Engineering} \\
\textit{Indian Institute of Technology}\\
Ropar, India \\
2016csb1059@iitrpr.ac.in }
\and
\IEEEauthorblockN{ Soumya Soubhagya Dandapat}
\IEEEauthorblockA{\textit{Computer Science and Engineering} \\
\textit{Indian Institute of Technology}\\
Ropar, India \\
2017csb1114@iitrpr.ac.in}

}

\maketitle

\begin{abstract}
   Advancement in  Processor technology has made it easy to handle data-intensive workloads, but limiting main memory advances has created performance bottlenecks. In DRAM, there have been improvements in DRAM access latency as well as reduction in cost-per-bit with the increase in cell density. But still DRAM data transfer rate lags behind the processing speed of the current generation processors. As Memory advancements based on hardware have been progressing at a slower pace, to cope up with High-end Processors,  Architectural level advancements such as Prediction techniques, Replacement policies, etc are the major subject. 

        In the recent field of research, Data prediction is a sought out topic as correct prediction can boost performance by decreasing the amount of excess memory access by predicting data beforehand using data access trends and behaviors. Though prediction techniques have been implemented at most of the Computer Architecture, We propose implementing data prediction in DRAM level architectures like TL-DRAM \cite{lee2013tiered} and CROW \cite{hassan2019crow}. Both of these method distributes the DRAM into different parts which contain a smaller section which is faster and larger section which contains the bulk of data but is comparatively slower. We wish to use data prediction in between these sections of memory to have predicted data transferred to the faster sections to improve the overall performance by reducing the memory access time.

\end{abstract}


\section{Introduction}
DRAM access latency is a challenge to improve system performance and energy efficiency. While DRAM capacity increased significantly over the last two decades, DRAM access latency decreased only slightly. The high latency of DRAM access degrades the performance of all other workloads. A lot of methods in the past have been proposed to decrease the access latency of DRAMs. Dividing the DRAM subarray into multiple segments for faster access of DRAM rows is one of the major contributions in reducing access latency of DRAM rows. These modifications achieved lower access latency for closer DRAM segments. Hence, reducing the overall access latency of DRAM. There was a significant improvement in access latency using these methods.
However, this architecture of faster and slower DRAM access introduces the drawback for applications whose working set lies in far segments. For such applications, accessing DRAM rows will lead to performance degradation due to slow access of far segments. 
This drawback takes us to the concept of prefetching which aims to boost application performance by fetching instructions or data to faster segments beforehand. This paper aims to apply the prefetching concept at DRAM level for such DRAM architectures. By using prefetching concepts and near segments as cache for far segments we aim to reduce performance degradation for such applications using these DRAM architectures.

% Advancement in processor technology make easy to handle data-intensive workload, but this causes performance bottlenecks in accessing the main memory. In DRAM cost-per-bit reduces with increased in cell density as well as improvement in DRAM access latency. With such advancement in main memory technology, DRAM data transfer rate still lag processing speed of speculating multi-core out-of-order processor.

% Eminent researcher contribute architectural level solution to over come this bottleneck, reducing the data moment between chip and main memory is one of them. 

% the data moment between different segment of DRAM

Our work is oriented to reduced such performance degradation by using prefetcher, efficient use of prefetcher help to reduce the compulsory misses. Theoretically we analysis different prefetcher (i.e. stride stream temporal and spatial prefetchers) and find their merit \& demerits. With all this profiling spatial prefetcher have advantage of recording footprint associated with pages with some pattern and is lookup multiple times with different event length.

Our proposal is to divide the DRAM rows into multiple segment and use concept of BINGO spatial prefetcher to reduce the inter-segment data migration latency. The concept is divide into two phases:
\begin{itemize}
\item In first phase we try to prefetch accurate data by using long event and provide to LLC. 
\item In second phase we try expand miss-coverage by using short event and try to provide inter row migration smoothly.   
\end{itemize}

Complete work is divided into the following sections: Related Work, Proposed Work, Tools for Experiments.
  
\section{Related Work}
As per knowledge, this was the first proposed in which prefetcher are used in different segment of DRAM for large data-intensive workload. Related work provide brief study over DRAM architecture and prefetcher.
To handle huge amounts of data in DRAM causes access latency issue in terms of tRCD (row to column delay) and tRC(refresh delay)\cite{jang2016refresh}.

\begin{figure}[ht]
\centering
\includegraphics[width=8cm]{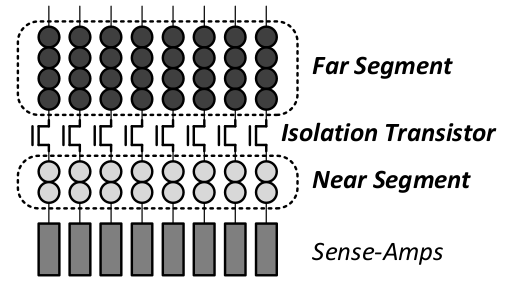}
\label{fig:tldram}
\caption{TL-DRAM architecture: Near Vs. Far Segments}
\end{figure}

TL-DRAM\cite{lee2013tiered} introduces the concept of reducing bit line length by using isolation transistors to decrease the bit line parasitic capacitance(directly proportional to length), which reduces the time tRCD for near segment and hence reduces the access latency of near segment of DRAM. They divided the DRAM subarray in two parts referred to as near(close to Sense Amplifiers) and far(remaining) segments. The near segment is directly connected to Sense Amplifiers and the far segment is connected to Sense Amplifiers using isolation transistors. Figure [\ref{fig:tldram}] shows the TL-DRAM architecture.

\begin{figure}[ht]
\centering
\includegraphics[width=8cm]{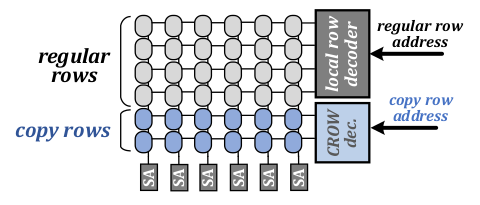}
\label{fig:crow}
\caption{CROW architecture: Regular and Copy rows in subarray}
\end{figure}

CROW\cite{hassan2019crow} introduces the concept of copy rows in DRAM subarrays. They divided the DRAM rows into two parts first representing regular rows and second representing copy rows. Some of these frequently accessed regular rows are copied to copy rows(for example let takes Row R1 is copied to Copy Row CR1), so that next time when row R1 is accessed both R1 and CR1 rows are activated in order to double the charge to drive the system and hence decreasing the time to reach stable state(read state) of sense amplifiers. Thus reducing the tRC and therefore reducing the access latency of DRAM for some rows. Figure [\ref{fig:crow}] shows the CROW architecture. 

In TL-DRAM and CROW, major performance degeneration is due to migration of row content in different row segments. Our work introduces prefetcher for mitigation of such a high migration latency. Author's proposed different type of prefetcher according to workload behaviour i.e. stride, streaming, temporal and spatial.
For strides detection Baer et al.\cite{10.1145/125826.125932} uses buffer of recent PC which was lookup by current PC for prediction of future upcoming addresses by it is not aggressive. Stream prefetcher\cite{palacharla1994evaluating} is aggressive in nature because stream buffer provide capability to track multiple stream of different access pattern simultaneously. Temporal prefetcher\cite{bakhshalipour2018domino} provides enough coverage but it requires a good amount of storage for storing the page level correlation and lookup mechanism that might affect the timeliness. BINGO\cite{8675188} a spatial prefetcher uses less storage for history table and try to make multiple lookup for different length of event. For better accuracy BINGO used long event and for good coverage lookup performed by shorter history.

\section{Proposed Work}
The main concept of this work is to improve the access latency of DRAM cells. In TL-DRAM\cite{lee2013tiered}, D. Lee et. al. tries to improve the access latency of DRAM cells by reducing the tRCD time of DRAM cell access. This is achieved by dividing the sub-array into two parts using isolation transistors. Similarly, in CROW\cite{hassan2019crow}, H. Hassan et. al aims to improve the access latency by using copy rows. The near segment also helps in reducing the access latency of frequently accessed cells of the far segment by transferring/copying the data from far segment to near segment. However, for good performance the number of DRAM cells in near-segments in both cases is limited, which leads us to the one disadvantage i.e. only few cells can have low access latency. Due to this limited size of near segments, there will always be a miss in the near segment for first access of memory addresses mapped in the far segment. For applications having working size greater than near segment size, there will be a miss in near segment for each memory access mapped in far segment. 
This paper proposes to overcome this problem by adding a prefetcher at DRAM level to avoid such misses by prefetching the far segment’s data to the near segment based on some prediction. Based on the data pattern, the prefetcher will decide which data to be prefetched from far segments to near segments. By adding prefetcher we can fetch the required before it's actual request which will reduce the overall access latency of DRAM. This won't reduce the actual tRCD or tRC of far or near segments but as now the data is already fetched in near segments, the overall access time as seen/observed by the processor will be reduced hence will reduce the stall time of processor to fetch data from memory. 

In case of TL-DRAM\cite{lee2013tiered}, near segment is used as LRU cache for far segments for which every first access to a data in far subarray will be a miss in near subarray. It can also be observed that the tRC of far segments in TL-DRAM is higher than the normal single subarray DRAM. This increase in tRC can lead performance degradation for those applications which continuously access the different rows of far segments. To overcome this problem they used near segments as hardware maintained LRU cache for far segments. However, as the author mentioned that to get benefit from this architecture the size of the near subarray should be small, which is one of main reasons of thrashing problems in cache. Also for such thrashing applications as each access is from a different row of far segment, the overall access latency will shoot up due to the high value of tRC. Thus leading to performance degradation of such applications. 

\begin{figure}[ht]
\centering
\includegraphics[width=8cm]{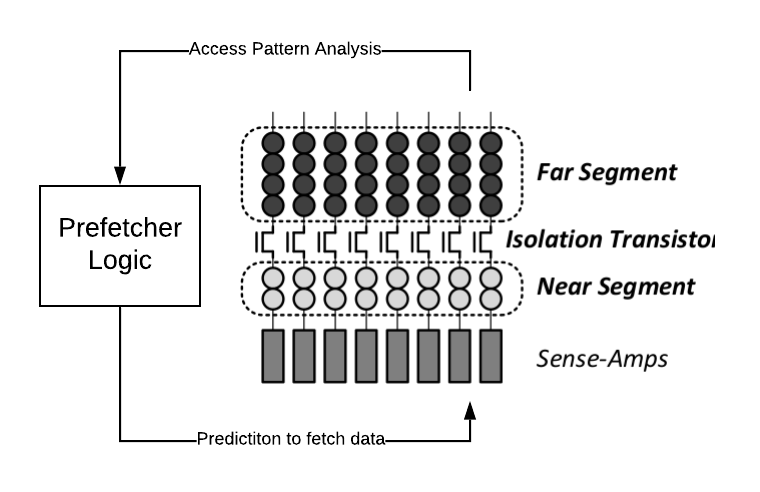}
\label{fig:memoryflow}
\caption{Proposed architecture}
\end{figure}

To solve this high access latency issue we propose to use a prefetcher that fetches the required data to near segments of DRAM subarray before time based on prediction by analysing the memory access pattern. Figure [\ref{fig:memoryflow}] shows the basic flow of proposed idea. With the help of prefetcher the data will already present in near subarray leading to a hit for most first access of far segment subarray and as the access latency of near segment is less than the access latency of normal single subarray DRAM architecture, the total access latency of each DRAM access will reduce. Also as most of the access will be a hit in near segments the overall access latency as observed by the processor will reduce. Hence, increasing the overall performance of applications.
Similarly for CROW\cite{hassan2019crow}, in order to utilize the copy rows in a more effective way we can add a prefetcher at DRAM level that prefetches the data from other rows to copy rows and hence decreasing the overall access latency of CROW DRAM architecture.

While implementing this architecture one important question arises i.e. Will there be an increase in refresh operations by twice for those data that are present at two places(one in copy/near segment and other in regular/far segment)? This is an important question that needs to be handled as DRAM refresh operation falls under one of the important challenges of improving DRAM. 
In this architecture, the increase in DRAM refresh operation will be very much comparable to CROW or TL-DRAM architecture, as we are only prefetching the required rows to copy rows, and as observed by authors of CROW paper that the number of hit rate of copy rows is very high therefore the reads and writes in copy rows will be very frequent and thus the refresh operations for these rows can be avoided as suggested by this paper. Hence, the refresh overhead will be comparable to CROW or TL-DRAM architecture. 

Prefetcher can be a hardware or software logic that uses program behaviour to boost its performance by fetching data to faster segments beforehand.
Adding a prefetcher in an architecture introduces some new challenges to be handled in order to achieve a successful or applicable architecture. Some of the main challenges that are faced in order to maximize  the  positive  performance  and  minimize  negative are mentioned below:
\begin{itemize}
    \item Where to add prefetcher logic ?
    \item How  to  generate a prefetch  history  buffer,  how the pattern  of page  co-relation  is  generated  and  what  and  where  the miss counter is placed and what is the size of MSHR?
    \item Addition of new commands to trigger the prefetcher functions like when to prefetch?
    \item Finally, effectiveness of prefetcher is evaluated in terms of:
    \begin{itemize}
        \item Prefetcher Accuracy: aggressiveness is directly proportional to prefetch accuracy.
        \item Prefetch Lateness: lateness is defined as ration of number of late prefetcher and no of useful, performance of accurate perfetcher is not increase if it is late.
        \item Cache Pollination: when prefetch request is flooded that causes no spaces of demand request this directly impact the performance. 
    \end{itemize}
\end{itemize}

\section{Tools for experiments}
For simulation and experimental analysis of this idea we decided to use either Ramutor \cite{kim2015ramulator} or DRAMsim \cite{wang2005dramsim}. We aimed for these simulators as both these are publicly available as well as these are widely used for both industrial and academic purposes. Ramulator \cite{kim2015ramulator} is a fast and cycle-accurate DRAM simulator that is designed by keeping in mind the idea of extensibility. Ramulator is a generalized template for modeling DRAM systems. It also provides support for a wide range of DRAM standards. Also, one of the important qualities of Ramulator is that it does not sacrifice simulation speed to gain extensibility. It has an architecture setup for TL-DRAM. Also, as authors of CROW paper have used it for their implementation(which is widely available), it becomes easy to add our manipulations directly into the Ramulator based CROW architecture. Second is DRAMSim2\cite{wang2005dramsim}. It is a detailed and C-based memory system simulator. It implements timing models for various existing memories including SDRAM, DDR2, etc. It allows the user to easily vary it’s parameters. DRAMsim also models the  power consumption of SDRAM and its derivatives. It can be used either standalone or can be integrated with other simulators for full system simulations. 
Finally we are aiming to use ChampSim\cite{ChampSim}, a trace-based simulator for a microarchitecture study, to analysis various prefetchers and find the best one that can be added efficiently to our architecture. 

\bibliographystyle{ieeetr}
\bibliography{Memory}

\end{document}